%% file: main.tex
\documentclass[journal]{IEEEtran}

\usepackage{amsmath,amsfonts,amssymb}
\usepackage{pstricks}
\usepackage{graphicx}
\usepackage{cite}
\usepackage{epstopdf}
\usepackage{array}
\usepackage{afterpage}
\usepackage{booktabs}
\usepackage{boldline} 
\usepackage{multirow}
\usepackage{float}

\usepackage{tikz}
\usetikzlibrary{patterns}
\usepackage{tabu}
\usepackage[subrefformat=parens,labelformat=empty,caption=false]{subfig}
\usepackage{stfloats}
\usepackage{enumitem}
\usepackage{color, colortbl}

\usepackage{algorithm}
\usepackage{algorithmic}

\input{mymacros} 
\graphicspath{{Figures/}{bioPhotos/}}

\ifCLASSINFOpdf
  
\else
  
\fi

\usepackage{tikz}
\usetikzlibrary{calc}
\usepackage{atbegshi}

\newcommand{\SubmissionNoticeOverlay}{%
  \begin{tikzpicture}[remember picture,overlay]
    \node[anchor=north, yshift=-8mm] at (current page.north) {%
      \parbox{0.97\paperwidth}{\centering\small
      This work has been submitted to the IEEE for possible publication.\\
      Copyright may be transferred without notice, after which this version may no longer be accessible.}%
    };
  \end{tikzpicture}%
}

\AtBeginShipoutFirst{\SubmissionNoticeOverlay}
\begin{document}
%

\title{Deep Learning-Driven Black-Box Doherty Power Amplifier with Pixelated Output Combiner and \\ Extended Efficiency Range}


\author{Han~Zhou,~\IEEEmembership{Member,~IEEE}, Haojie~Chang~\IEEEmembership{Member,~IEEE}, and David~Widén
    \thanks{This research was supported in part by Swedish Innovation Agency (Vinnova) Grant 2024-02531, MULTIRACS, through the Eureka CELTIC Framework, and in part by VINNOVA, Sivers Semiconductors, and Chalmers University of Technology under Grant 2022-00863.}
    \thanks{H.~Zhou, H.~Chang, and D.~Widén are with Chalmers University of Technology, Sweden (e-mail: han.zhou@ieee.org).} 
}

\maketitle


\begin{abstract}
This article presents a deep learning-driven inverse design methodology for Doherty power amplifiers (PA) with multi-port pixelated output combiner networks. A deep convolutional neural network (CNN) is developed and trained as an electromagnetic (EM) surrogate model to accurately and rapidly predict the S-parameters of pixelated passive networks. By leveraging the CNN-based surrogate model within a black-box Doherty framework and a genetic algorithm (GA)-based optimizer, we effectively synthesize complex Doherty combiners that enable an extended back-off efficiency range using fully symmetrical devices. As a proof of concept, we designed and fabricated two Doherty PA prototypes incorporating three-port pixelated combiners, implemented with GaN HEMT transistors. In measurements, both prototypes demonstrate a maximum drain efficiency exceeding $74\%$ and deliver an output power surpassing $44.1~\mathrm{dBm}$ at $2.75~\mathrm{GHz}$. Furthermore, a measured drain efficiency above $52\%$ is maintained at the $9$-dB back-off power level for both prototypes at the same frequency. To evaluate linearity and efficiency under realistic signal conditions, both prototypes are tested using a $20$-MHz 5G new radio (NR)-like waveform exhibiting a peak-to-average power ratio (PAPR) of $9.0$-dB. After applying digital predistortion (DPD), each design achieves an average power-added efficiency (PAE) above $51\%$, while maintaining an adjacent channel leakage ratio (ACLR) better than $-60.8~\mathrm{dBc}$.

 
\end{abstract}



%
\IEEEpeerreviewmaketitle

\section{Introduction}
%
%
%
%

\IEEEPARstart{T}o meet the increasing demand for higher data rates, modern communication systems have embraced advanced modulation techniques. These methods, however, inherently produce signals with a high peak-to-average power ratio (PAPR). Meanwhile, energy efficiency has become a critical design consideration, as power consumption in mobile networks constitutes a substantial share of both operational costs and environmental impact. Excessive energy usage further complicates system design due to the need for additional cooling infrastructure to manage thermal dissipation. Among all radio front-end components, power amplifiers (PA) are typically the most power-intensive. Therefore, improving PA efficiency, particularly under backed-off output power conditions, can yield considerable energy savings and lead to more sustainable and cost-effective network deployments~\cite{HW_CF}.

Active load-modulated PA architectures have gained considerable attention for their ability to enhance efficiency at back-off power levels. Recent innovations in this area~\cite{LMBA, SLMBA_TCAS1, SLMBA_TCAS2, DEPA_1, DEPA_TCAS1, NoLMPA, CLMA_Han2} have demonstrated notable improvements in back-off efficiency. However, their circuit complexity often hinders practical deployment, especially at higher frequencies. In contrast, the Doherty PA~\cite{DohertyPA} exhibits a simpler circuit configuration, contributing to its widespread commercial adoption. Central to Doherty PA performance is the output combiner network, which govern the interaction between the main and auxiliary sub-amplifiers. This cooperative behavior directly impacts achievable back-off efficiency.

In parallel, inverse design methodologies have emerged as a powerful paradigm across diverse fields such as integrated photonics~\cite{Tahersima2019DeepNN, Ma2020DeepLF, Molesky2018InverseDI}, meta-lens antennas~\cite{DL_Meta1, DL_Meta2, DL_Meta3}, and scattering theory~\cite{DL_S1, DL_S2}. The core principle is to define a target performance and then determine the optimal structure through a top-down approach. This often involves pixelating a designated layout area into an $N \times M$ binary matrix, where each pixel denotes the presence or absence of a specific material. Compared to conventional electromagnetic (EM) structures composed of lumped or distributed elements, inverse design enables exploration of a vastly larger design space. In recent years, deep learning has been increasingly applied to such problems for its ability to learn complex features and efficiently navigate high-dimensional design spaces~\cite{lecun2015deep}.

In radio frequency (RF) circuit design~\cite{DL_Microwave}, deep learning–based inverse design has been successfully applied to broadband PAs, passive circuits~\cite{DL_PA1, DL_PA2, DL_PA3, AI_sensen, AIIMS_chu}, and harmonic-tuned Class~F PAs~\cite{AI_HZ}. However, most prior work has focused on relatively standard structures such as power splitters, filters, and Class~B PAs with conventional combiners. Doherty PAs, by contrast, require much more sophisticated three-port combiner networks that integrate load modulation, power combining, phase shifting, and impedance transformation within a compact footprint.

To address this gap, we extend the inverse design methodology to the synthesis of complex pixelated three-port Doherty combiner networks. Inverse design of a traditional Doherty combiner is particularly challenging because it is typically developed through an iterative process, resulting in multiple cascaded networks such as matching sections, parasitic compensation stages, offset lines, and post-matching circuits in addition to the load modulation network~\cite{DPA_TCAS1, DPA_JSSC, DPA_TCAS2, DPA_TCAS3}. This complexity significantly limits the practicality of applying inverse design directly to conventional combiners. By contrast, adopting the black-box design approach of~\cite{Dohertyblackbox, DohertyHan} allows direct synthesis of the complete combiner from transistor load-pull data, eliminating the need to separately design and integrate each sub-network. This methodology greatly smooths the design process and enables extended efficiency enhancement within fully symmetrical configurations. Nevertheless, existing black-box approaches are typically limited to combiner structures composed of lumped or distributed circuit elements, such as transmission lines, capacitors, and inductors. These methods rely on bottom-up design strategies that involve parameter sweeps and repeated EM simulations. Although such techniques provide insight into circuit behavior, they inherently constrain the design space to predefined topologies with limited flexibility and a small number of tunable parameters. As a result, existing methodologies restrict the discovery of higher‑performance solutions, require considerable simulation and optimization time, and ultimately impede the exploitation of the full performance potential of Doherty combiner networks.

\begin{figure*}[t!]
    \centering   
    \includegraphics[width=\textwidth]{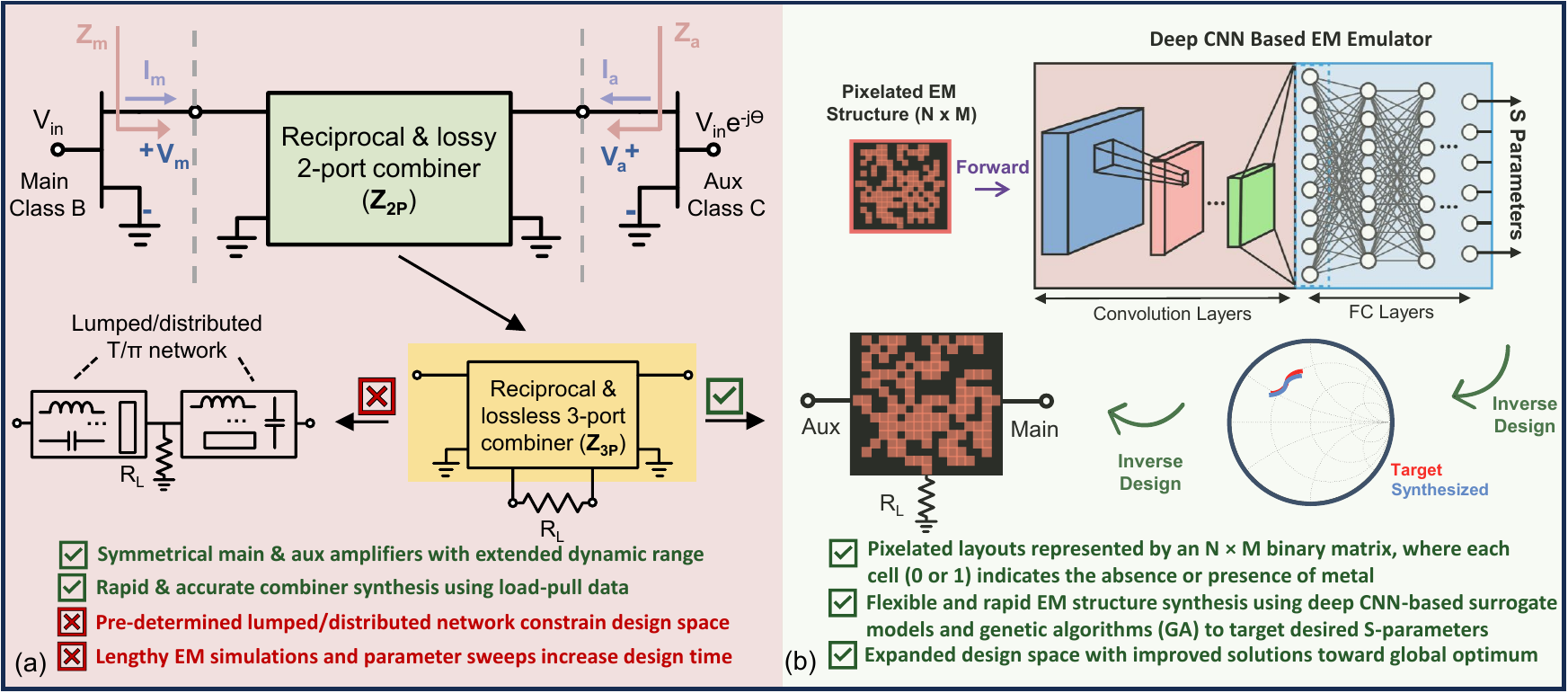}
    \caption{Overview of the proposed deep learning–driven approach for the synthesis and design of black-box Doherty PAs with symmetrical size and extended back-off efficiency range. (a) Analytical black-box approach for deriving combiner parameters from load-pull data. (b) Deep learning–based top-down approach employing a pixelated Doherty combiner layout to explore the full design space.}
    \label{fig:intro}
\end{figure*}

Motivated by these limitations, this work introduces a data‑driven inverse design framework that relaxes the topological constraints and mitigates the dependence on time‑consuming EM simulations and tuning in existing Doherty combiner synthesis approaches. Building on this foundation, we present the first Doherty PA design that combines deep learning–driven inverse design with a pixelated output combiner network (Fig.~\ref{fig:intro}). A deep convolutional neural network (CNN) surrogate model is trained to map pixelated EM structures to their scattering parameters (S-parameters) with high accuracy. Coupled with a genetic algorithm (GA), this surrogate accelerates the inverse design process by eliminating repeated full-wave EM simulations. Integrated with the black-box Doherty framework, our method achieves significant efficiency improvements at extended back-off levels using identical transistors directly synthesized from their load-pull data. The remainder of this paper is organized as follows. Section~II illustrates the black-box combiner synthesis methodology and analyzes the Class B/C Doherty PA. Section~III details the CNN architecture, data generation workflow, and GA strategy. Section~IV demonstrates the implementation process, including two Doherty PA prototypes. Section~V reports measurement results. Section~VI concludes the paper.

\section{The Black-box Combiner Theory}
\label{Sec2}
In this section, the generalized combiner synthesis approach is presented. The parameters of a generic two-port combiner are first derived to analyze the Class B/C Doherty PA behavior with extended efficiency range under varying drive conditions. Then, the synthesis procedure of a two-port Doherty combiner using load-pull data is demonstrated, along with the condition required to transform it into a lossless three-port combiner with resistive output loading.

\subsection{Black-Box Doherty PAs}
Following the analysis in~\cite{DohertyHan}, transistors are modeled as ideal, piecewise, voltage-controlled current sources. Only the fundamental component is considered, with higher harmonics short-circuited, representing ideal Class B operation. A phase delay $\theta$ is assumed between the output currents of the main and auxiliary amplifiers. The output currents are given by~\cite{CLMA_Han2}
\begin{equation}
    I_{\text{m}} = \beta i_{\text{m, M}}
\end{equation}
\begin{equation}
    I_{\text{a}} = \begin{cases} \displaystyle 0, & 0\leq  \beta \leq \beta_{\text{B}} \\ \displaystyle \left(\frac{\beta - \beta_{\text{B}}}{1 - \beta_{\text{B}}}\right)i_{\text{a,M}}\cdot e^{-j\theta}, & \beta_{\text{B}} \leq  \beta \leq 1 
    \end{cases}
\end{equation}
where $\beta$ and $i$ denotes the normalized input voltage and fundamental current, respectively. In these expressions, the first subscript (m or a) refers to
the main or auxiliary amplifier, respectively, and the second
subscript (M or B) denotes operation at maximum or back-off
power levels.

The combiner network is described by its impedance matrix $\mathbf{Z}_{\text{2P}}$, relating voltages and currents at the amplifier outputs

\begin{gather}
\label{eqn.VI}
 \begin{bmatrix} V_{\text{m}} \\ V_{\text{a}} \end{bmatrix}
 =\mathbf{Z_{\text{2P}}}\begin{bmatrix}
   I_{\text{m}} \\
   I_{\text{a}}
   \end{bmatrix} =
  \begin{bmatrix}
   Z_{11} & Z_{12}  \\
   Z_{12} & Z_{22}  
   \end{bmatrix}
   \begin{bmatrix}
   I_{\text{m}} \\
   I_{\text{a}}
   \end{bmatrix}
\end{gather}

\begin{figure}[t!]
    \centering         \includegraphics[width=0.8\columnwidth]{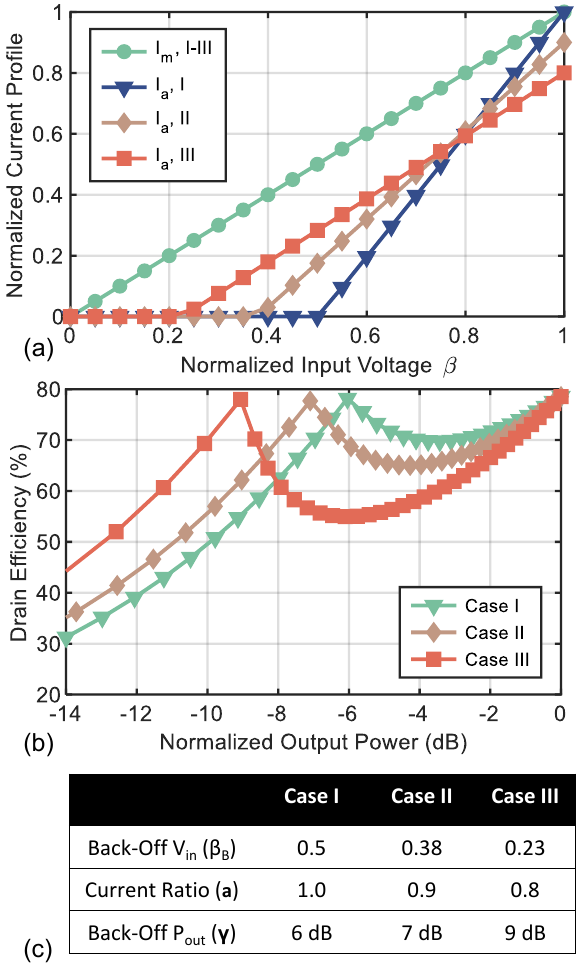}
    \caption{Illustration of the black-box technique under different driving conditions. The output power is normalized to the corresponding saturated output power for each current‑scaling ratio. (a) Normalized current drive profiles as a function of input voltage. (b) Theoretical performance comparison for different drive profiles (Cases I–III). (c) Corresponding current ratios and back-off levels for each case.}
    \label{fig:2}
\end{figure}

Assuming an optimal load of $R_{\text{opt}}$ for the main amplifier, the auxiliary amplifier’s optimal load reduces to $R_{\text{opt}}/\alpha$, with $\alpha$ defined as the ratio of peak output currents between the main and auxiliary amplifiers. The combiner’s impedance parameters are then derived as~\cite{zhou2023phd}

\begin{equation}
\left\{
\begin{aligned}
Z_{11} & = \displaystyle \frac{R_{\text{opt}}}{\beta_{\text{B}}}\\
Z_{12}  &= \left(1-\displaystyle\frac{1}{\beta_{\text{B}}}\right)\left(\frac{R_{\text{opt}}}{\alpha}\right)e^{j\theta}\\
Z_{22} & = \displaystyle \left(\frac{1}{\beta_{\text{B}}}+\alpha e^{-j2\theta}-1\right)\left(\frac{R_{\text{opt}}}{\alpha^2}\right)e^{j2\theta}
\end{aligned}
\right.
\label{eqn.Zideal}
\end{equation}

To transform the lossy reciprocal two-port network into an equivalent lossless three-port combiner with resistive termination, the following condition must be satisfied~\cite{Dohertyblackbox}
\begin{equation}
     \Re\{Z_{12}\}^{2} = \Re\{Z_{11}\}\Re\{Z_{22}\}
     \label{eqn.bc}
\end{equation}

The normalized output power back-off level, denoted by $\gamma$, and the phase delay $\theta$ are derived as~\cite{zhou2023phd}

\begin{equation}
\label{gamma}
     \gamma=\displaystyle \frac{1+\alpha}{\beta_{\text{B}}}
\end{equation}

\begin{equation}
\label{theta}
     \theta = k\pi \pm \displaystyle \arcsin{\sqrt{\frac{\beta_{\text{B}}\left(\alpha - \beta_{\text{B}} + 1\right)}{1-\beta_{\text{B}}^2}}}
\end{equation}
where $k$ is an arbitrary integer.

Expressions (\ref{eqn.Zideal}–\ref{theta}) indicate that, for a given back-off level $\gamma$ and current ratio $\alpha$, a valid phase shift $\theta$ and corresponding combiner impedance matrix $\mathbf{Z}_{\text{2P}}$ can be determined. As illustrated in Fig. 2, the theoretical analysis suggests that this method enables the realization of desired back-off levels across various current ratios, facilitating full utilization of the main and auxiliary transistors.

\subsection{Combiner Synthesized from Load-Pull Data}
The theoretical analysis above demonstrates the fundamental operating principle of the black-box approach. However, at high frequencies, transistor parasitics and nonlinearities significantly influence PA performance. As a result, a combiner design methodology guided by nonlinear device behavior under large-signal excitation becomes essential for fully utilizing the device capabilities. The two-port combiner impedance matrix ($\mathbf{Z}_{\text{2P}}$) can be derived from load-pull data, as follows.

\begin{equation}
\left\{
\begin{aligned}
Z_{11} & = \displaystyle \frac{\left(Z_{\mathrm{a,M}}+Z_{\mathrm{a,off}}\right)Z_{\mathrm{m,B}}\alpha^{2} + \left(Z_{\mathrm{m,M}}-Z_{\mathrm{m,B}}\right)Z_{\mathrm{m,M}}}{Z_{\mathrm{m,M}}-Z_{\mathrm{m,B}}+\left(Z_{\mathrm{a,M}}+Z_{\mathrm{a,off}}\right)\alpha^{2}}\\
Z_{12}  & = \displaystyle \frac{\left(Z_{\mathrm{m,M}}-Z_{\mathrm{m,B}}\right)\left(Z_{\mathrm{a,M}}+Z_{\mathrm{a,off}}\right)\alpha}{Z_{\mathrm{m,M}}-Z_{\mathrm{m,B}}+\left(Z_{\mathrm{a,M}}+Z_{\mathrm{a,off}}\right)\alpha^{2}}\\
Z_{22} & = \displaystyle \frac{\left(Z_{\mathrm{a,M}}+Z_{\mathrm{a,off}}\right)Z_{\mathrm{a,M}}\alpha^{2}+\left(Z_{\mathrm{m,B}}-Z_{\mathrm{m,M}}\right)Z_{\mathrm{a,off}}}{Z_{\mathrm{m,M}}-Z_{\mathrm{m,B}}+\left(Z_{\mathrm{a,M}}+Z_{\mathrm{a,off}}\right)\alpha^{2}}
\end{aligned}
\right.
\label{eqn.Zcomplex}
\end{equation}

The parameter $\alpha$ is given by

\begin{equation}
    \alpha = \displaystyle \frac{i_{\text{a,M}}}{i_{\text{m,M}}}=\sqrt{\frac{\Re\{Z_{\text{m,M}}\}P_{\text{a,M}}}{\Re\{Z_{\text{a,M}}\}P_{\text{m,M}}}}e^{-j\theta}
    \label{eqn.alpha}
\end{equation}
where $P$ represents the delivered output power and $Z$ is the optimal impedance, both from load-pull data.

Furthermore, when selecting the design parameter for the back-off power level ($\gamma_{\mathrm{B}}$), the following relationship must be satisfied to ensure power conservation
\begin{equation}
    \displaystyle \gamma_{\mathrm{B}} P_{\text{m,B}} = P_{\text{m,M}} + P_{\text{a,M}}
    \label{eqn.pc}
\end{equation}

To implement a physically realizable, lossless three-port combiner ($\mathbf{Z}_{\text{3P}}$) terminated with a resistive load $R_L$, the synthesized lossy two-port network ($\mathbf{Z}_{\text{2P}}$) must satisfy the condition described in equation (5). Given equations (\ref{eqn.bc}), (\ref{eqn.Zcomplex}), and (\ref{eqn.alpha}), the only remaining unknown is the phase parameter $\theta$, which can be solved analytically or numerically. Once $\theta$ is determined, the complete impedance matrix $\mathbf{Z}_{\text{2P}}$ for the combiner can be fully specified.


\begin{figure*}[t!]
    \centering   
    \includegraphics[width=\textwidth]{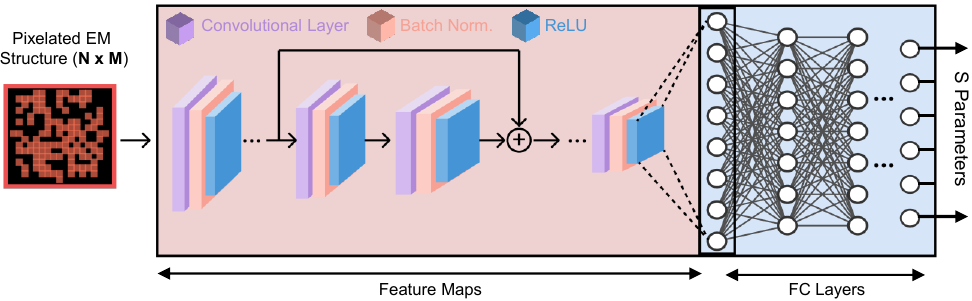}
    \caption{Architecture of the trained deep CNN with residual connections. The input is a binary $15 \times 15$ matrix representing the pixelated EM layout structure, and the output is the predicted real and imaginary components of the corresponding S-parameters across the frequency range of interest.}
    \label{fig:CNN}
\end{figure*}

\begin{figure}[t!]
    \centering   
    \includegraphics[width=0.618\linewidth]{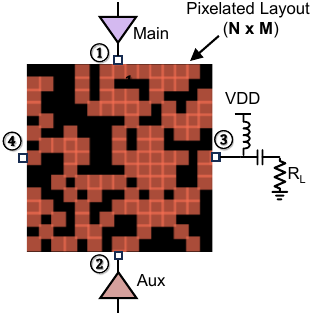}
    \caption{The employed pixelated Doherty combiner networks, where the feed locations for the main (port~$1$), auxiliary (port~$2$), and output (port~$3$) are placed at the center of each corresponding edge of the $15$-pixel array. The output port~$3$ is connected to the load ($R_{\mathrm{L}}$), while port~$4$ is left open.}
    \label{fig:combiner}
\end{figure}

\section{Deep Learning-Driven Inverse Synthesis of Doherty Combiner Network}
\label{Sec3}
Building upon the theoretical analysis and equations in Section~\ref{Sec2}, a two-port Doherty combiner matrix $\mathbf{Z}_{\text{2P}}$ can be directly derived from load-pull data, enabling designs with extended back-off efficiency while fully utilizing the employed transistors. However, physical realizations of such combiners have traditionally relied on predefined lumped or distributed topologies, often depending on expert intuition, iterative EM simulations, and manual parameter tuning~\cite{Dohertyblackbox, DohertyHan}.

\begin{table}[!t]
\renewcommand{\arraystretch}{1.2}
\caption{Detailed Configuration of the deep CNN Architecture}
\label{tab:CNN}
\centering
\begin{tabular}{@{}cccc@{}}
\toprule
\textbf{Conv. Layer} & \textbf{Filter Size} & \textbf{Filter Number} & \textbf{Inputs} \\ 
\midrule
$C_1$  & $12 \times 12$ & 32  & Matrix\textsubscript{in}                                                         \\
$C_2$  & $12 \times 12$ & 32  & $C_1$                  \\
$C_3$ & $10 \times 10$ & 32  &  $C_2$ + Matrix\textsubscript{in} \\
$C_4$  & $10 \times 10$ & 32  & $C_3$                  \\
$C_5$  & $8 \times 8$   & 32  & $C_4$ + $C_2$          \\
$C_6$  & $8 \times 8$   & 32  & $C_5$                 \\
$C_7$  & $6 \times 6$   & 32  & $C_6$ + $C_4$          \\
$C_8$  & $6 \times 6$   & 32  & $C_7$                 \\
$C_9$  & $4 \times 4$   & 32  & $C_8$ + $C_6$         \\
$C_{10}$ & $4 \times 4$   & 32  & $C_9$                 \\
$C_{11}$ & $3 \times 3$   & 32  & $C_{10}$ + $C_{8}$         \\
$C_{12}$ & $3 \times 3$   & 32  & $C_{11}$                 \\
\midrule
\textbf{FC Layer} & \textbf{Num. Neurons} & \textbf{Dropout} & \textbf{Act. Func.} \\
\midrule
$F_1$ & 2048 & 0.25 & LeReLU (0.01) \\
$F_2$ & 2048 & 0.25 & LeReLU (0.01) \\
$F_3$ & 2048 & 0.25 & LeReLU (0.01) \\
$F_4$ & 2048 & 0.25 & LeReLU (0.01) \\
$F_5$ & 2048 & 0.25 & LeReLU (0.01) \\
$F_6$ & 78   & 0    & Tanh          \\
\bottomrule
\end{tabular}
\end{table}

To overcome these limitations, this section introduces an inverse design methodology for synthesizing Doherty combiners using pixelated output networks. This approach enables exploration of a broader design space encompassing all fabricable structures. We begin by describing the pixelated layout design space and the corresponding training dataset. This is followed by the development of a deep CNN used as an EM surrogate model, as shown in Fig.~\ref{fig:CNN}. Lastly, the inverse synthesis process is detailed using a GA optimizer.

\subsection{Pixelated Doherty Combiner Network}
To synthesize compact and low-loss Doherty combiner networks, a planar circuit layout is defined over a ground plane and discretized into a two-dimensional binary pixel grid. Each pixel is assigned a binary value: “1” represents the presence of metal, and “0” denotes a dielectric (non-metal) region. The layout is partitioned into a $15 \times 15$ grid, where each pixel measures $1.8~\mathrm{\times}~1.8~\mathrm{mm}$. This resolution is selected to balance EM coupling fidelity and the feasibility of model training~\cite{AI_HZ}. A higher pixel resolution enables finer structural detail and access to a larger design space, potentially yielding improved performance at the expense of increased computational requirements. To ensure reliable electrical connectivity, particularly across diagonal paths, a $20\%$ metal overlap is applied to each metal pixel. This results in a total physical combiner size of $27.36~\mathrm{mm} \times 27.36~\mathrm{mm}$. 

As illustrated in Fig.~\ref{fig:combiner}, the feed locations for the main (port~$1$) and auxiliary (port~$2$) amplifiers, as well as the output (port~$3$) connected to the load resistance ($R_{\mathrm{L}}$), are placed at the center of each corresponding edge of the 15-pixel array. To enable DC biasing of the transistors, supply feeds (VDD) are introduced ahead of $R_{\mathrm{L}}$, allowing the implementation of RF chokes and DC blocks. 

The binary representation of this pixelated layout results in a total design space of $2^{225}$ possible EM structures. Given the vastness of this space, brute-force exploration is computationally infeasible in both time and memory. This challenge presents a compelling opportunity for deep learning. If a deep neural network can be trained as a surrogate EM model using a much smaller subset of simulation data ($<100000$), it can then accurately and rapidly predict the scattering parameters of arbitrary pixelated layouts. This surrogate model effectively replaces time-intensive EM simulations governed by Maxwell’s equations. When combined with a GA optimizer and guided by a target specification, the inverse design of pixelated structures can be effectively achieved.

\subsection{Deep Convolutional Neural Network}
The pixelated circuit layout can be naturally interpreted as a structured image, making CNNs, widely used in computer vision and image processing, well suited for this task. CNNs effectively model local spatial interactions and capture EM coupling between adjacent pixels by learning hierarchical features. In the following texts, we present the dataset generation process used for CNN training and describe the architecture of the proposed deep CNN model.

\subsubsection{Data Generation and Augumentation}
A dataset comprising diverse binary matrices representing pixelated EM structures and their corresponding S-parameters is essential for training the CNN. We fully automated the data generation process using Python scripts, which create pixelated layouts in Keysight Advanced Design System (ADS) and execute EM simulations using Momentum in ADS. The training circuits are generated by assigning each layout a random percentage of metal pixels, drawn from a normal distribution centered at $50\%$ with a standard deviation of $15\%$. During data generation, we ensure that each circuit maintains connectivity between the ports by applying a depth-first search (DFS) algorithm to verify continuous electrical paths. We simulate a total of 77000 four-port pixelated EM structures, running four simulations in parallel. Each circuit simulation takes around 24 seconds. Using an Intel i5-14600KF CPU, the total time required to generate the dataset is approximately 129 hours.

Each simulated structure includes four ports, with one port terminated in an open condition and another port terminated with a $50~\Omega$ load during augmentation. By applying rotations, translations, and mirror reflections~\cite{DL_PA1}, we generate additional three-port EM structures with a $50~\Omega$ load termination, where the S-parameters are derived directly from the original four-port simulations. We augment the data such that for every simulated circuit, eight training data points are created, significantly reducing the effective dataset creation time. After augmentation, the total dataset size is approximately $1.5~\mathrm{GB}$.

\begin{algorithm}[!t]
\caption{Genetic Algorithm Optimization}
\label{alg:GA_synthesis}
\begin{algorithmic}[1]
\REQUIRE Desired S-parameter profile; trained CNN surrogate model; population size $N$; maximum iterations $X$ etc.
\ENSURE Optimal pixelated layouts with target S-parameters.
\STATE Select a target S-parameter profile.
\STATE Generate $N$ binary matrices with direct connections, sampled from a normal metal pixel density distribution.
\FOR{each generation up to $X$ iterations}
    \STATE Predict S-parameters of all candidate circuits using the trained deep CNN model.
    \STATE Calculate fitness based on deviation from the target S-parameter profile.
    \IF{fitness of best candidate meets target}
        \STATE \textbf{break} loop; optimal solution found.
    \ENDIF
    \STATE Retain the top 10 circuits as elite individuals for the next generation.
    \STATE Add up to $30\%$ new random circuits depending on current iteration count to maintain diversity.
    \STATE Perform tournament selection:
        \begin{itemize}
            \item Split and recombine rows for crossover.
            \item Flip $1{-}10\%$ of pixels randomly for mutation.
            \item Fill population up to $N$ circuits.
        \end{itemize}
\ENDFOR
\RETURN The best pixelated circuit layout achieving or approaching the target S-parameter profile.
\end{algorithmic}
\end{algorithm}

\subsubsection{Deep CNN Architecture}
As illustrated in Fig.~\ref{fig:CNN}, the deep CNN architecture consists of twelve convolutional layers and six fully connected (FC) layers. A residual network (ResNet) structure is employed to mitigate the vanishing gradient problem, which improves training stability and accelerates convergence~\cite{He_2016_CVPR}. Leaky rectified linear unit (LeReLU) functions are applied as activation functions after batch normalization~\cite{ReLU}, enabling the network to learn complex nonlinear relationships. In the convolutional layers, local spatial features are effectively extracted from the binary input matrix representing the pixelated EM layout. These feature maps are then flattened into a one-dimensional vector, which is passed to the FC layers for further processing. During training, dropout layers are inserted to reduce the risk of overfitting~\cite{Dropout}. Each convolutional layer uses $32$ filters, and each FC layer contains $2048$ neurons with a dropout rate of $25\%$. The CNN takes binary matrices as input and produces real and imaginary components of the S-parameters as output. A detailed overview of the network architecture is provided in Table~\ref{tab:CNN}.

We train the network using the mean absolute error (MAE) loss function, which quantifies the average absolute difference between predicted and actual values, and is defined by
\begin{equation}
L = \frac{1}{N} \sum_{i=1}^{N} \left| y_i - \hat{y}_i \right|
\end{equation}
where $y_i$ and $\hat{y}_i$ denote the actual and predicted values, respectively. The Adam optimizer~\cite{adam} is used with an initial learning rate of $0.001$, which we reduce by a factor of $0.93$ every $10$ epochs to aid convergence. The neural network is trained for $300$ epochs using a batch size of $2790$. Upon completion, the model achieves a mean absolute error (MAE) of $0.068$, demonstrating high prediction accuracy.


\subsection{Genetic Algorithm-Based Inverse Synthesis}
The inverse synthesis of pixelated Doherty combiners is performed using a GA to efficiently explore the vast design space. As illustrated in Algorithm~\ref{alg:GA_synthesis}, the process begins by defining a desired S-parameter profile, which serves as the target specification. We then generate an initial population of $N=4000$ binary matrices representing connected circuit layouts, sampling metal pixel densities from a normal distribution to ensure a diverse starting point. 

\begin{table*}[!t]
\centering
\footnotesize
\caption{Load-pull Data Obtained at $2.75~\mathrm{GHz}$}
\label{tab:loadpull}
\begin{tabular}{|c|c|c|c|c|c|c|c|}
\hline
 & \multicolumn{3}{c|}{Main Amplifier} & \multicolumn{3}{c|}{Auxilary Amplifier} \\
\hline
 & $Z_{\mathrm{opt}}$ ($\Omega$) & $P_{\mathrm{del}}$ (dBm) & PAE ($\%$) & $Z_{\mathrm{opt}}$ ($\Omega$) & $P_{\mathrm{del}}$ (dBm) & PAE ($\%$) \\
\hline
Peak Power Level & $14.3+j1.6$ & $42.7$ & $74$ & $14.3+j1.6$ & $42.1$ & $78$\\
\hline
Back-Off Power Level & $7.2+j15.3$ & $36.4$ & $57$ & $0.25+j21.1$ (Off) & - & - \\
\hline
\end{tabular}
\end{table*}

The trained CNN surrogate model is used to rapidly predict the S-parameters of all candidate circuits in each generation, and fitness scores are calculated based on the deviation from the target S-parameter profile. If the fitness of the best candidate circuit meets the target performance or after $X=240$ iterations, the algorithm is terminated. Otherwise, the top $10$ circuits with the highest fitness scores are retained as elite individuals for the next generation. 

To maintain diversity, up to $30\%$ of the population is replaced with new random circuits depending on the current iteration count. This encourages exploration during the early stages of the optimization while focusing on refinement in later generations. Tournament selection is performed to identify parent pairs for crossover, where rows of parent matrices are split at random points and recombined to produce offspring. We also apply a mutation step by flipping between $1\%$ and $10\%$ of the pixels in the offspring circuits, introducing additional variability and helping to avoid convergence to local minima. This iterative process continues until the termination criterion is satisfied, gradually refining the population toward circuits that match the desired S-parameter response.


\section{Deep Learning-Driven Doherty PA Design}
In this section, we present the design of two prototype Doherty PAs using the proposed black-box synthesis methodology augmented by deep learning. The objective is to demonstrate the complete design procedure and validate the effectiveness of the proposed approach. 

To ensure a fair and focused evaluation, the input networks of both Doherty PAs are kept completely identical. The only difference lies in the output Doherty combiners, which are individually synthesized using the proposed deep learning-aided inverse design strategy. This setup highlights the diversity of synthesis solutions achievable with the method, as well as its robustness and flexibility.

For both the main and auxiliary branches, we employ the same commercially available 10-W GaN HEMT transistor form (CG2H40010F, MACOM) as the active device. This choice ensures symmetry in the device characteristics.

\subsection{Black-Box Doherty PA Design Based on Load-Pull Data}

We begin by designing both amplifier branches using the MACOM transistors (CG2H40010F). The input network, including matching, stabilization, and biasing circuits, is identically designed and fixed for both Doherty prototypes. Low‑impedance open stubs connected to the transistor gates, together with series RC networks at the inputs, are employed to simultaneously ensure unconditional stability and match the optimal source impedances of the transistors. In addition, resistors are incorporated in the gate bias lines to further enhance stability. A Wilkinson power divider is then used to equally distribute the input signal to the main and auxiliary amplifier branches. Next, we select the desired back-off power level ($\gamma_{\mathrm{B}} = 9~\mathrm{dB}$) and configure the gate bias of the main amplifier. Note that main amplifier’s gate bias involves a trade-off between gain and efficiency: biasing the main amplifier closer to Class~AB operation improves gain performance but can reduce efficiency. In our design, we set the main amplifier’s gate bias to achieve a quiescent current of $40~\mathrm{mA}$. For the auxiliary amplifier, we set the gate bias in Class~C operation. Selecting the auxiliary gate bias may require iterative adjustments to ensure that the auxiliary amplifier activates at the desired $9~\mathrm{dB}$ back-off level.

Load-pull characterization of the main and auxiliary amplifiers is performed at both the peak output and the specified back-off power level. These simulations are conducted using the bias conditions established in the preceding design phase. It should be noted that when selecting the optimal load impedance, multiple degrees of freedom exist to trade off efficiency, output power, and linearity. In this work, we prioritize maximizing efficiency at the back‑off power level and maximizing delivered output power at the peak power level. The extracted optimal impedances and corresponding power performance for both Doherty prototypes at peak and back-off conditions are summarized in Table~\ref{tab:loadpull}. Using equations (\ref{eqn.Zcomplex})–(\ref{eqn.pc}), together with equation (\ref{eqn.bc}), we derive the phase parameter solution of $\theta=133.4^{\circ}$ and the corresponding two-port combiner impedance matrix ($\mathbf{Z}_{\text{2P}}$) as
\begin{gather}
 \mathbf{Z_{2P}} 
 =
  \begin{bmatrix}
  $1.35$\;+\;j$6.94$\, & -$5.37$\;+\;j$14.02$  \\
   -$5.37$\;+\;j$14.02$ & $21.27$\;+\;j$16.10$  
   \end{bmatrix}
\end{gather}

The obtained $\mathbf{Z}_{\text{2P}}$ matrix fully specifies the desired Doherty combiner behavior in a reciprocal two-port black-box form, specifying the impedance relationships required for achieving proper load modulation and matching. By implementing this impedance matrix as a lossless three-port combiner with all lossy resistive components consolidated into a single port, the Doherty PA can achieve the correct load trajectories across the entire back-off range, enabling an efficiency enhancement at $\gamma_{\mathrm{B}}=9~\mathrm{dB}$. Successfully implementing this combiner network represents a critical step, as the associated design space is vast and requires systematic exploration.

\begin{figure*}[!t]
    \centering   
    \includegraphics[width=0.95\linewidth]{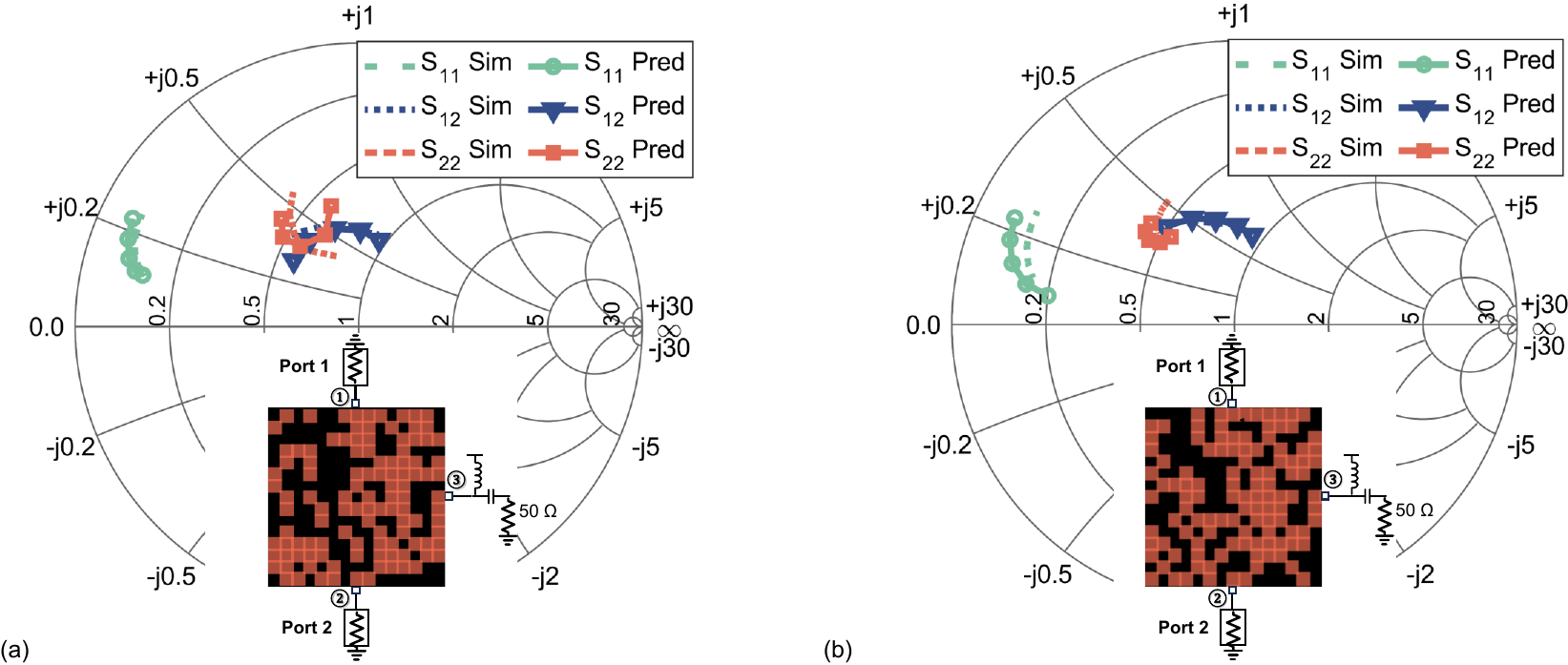}
    \caption{The two synthesized pixelated Doherty combiner networks, (a) and (b), along with their EM-simulated S-parameter results compared to the corresponding responses predicted by the deep learning approach, over the frequency range of $2.55$–$2.95~\mathrm{GHz}$.}
    \label{fig:sim_vs_pred}
\end{figure*}
\begin{figure*}[!t]
    \centering   
    \includegraphics[width=\textwidth]{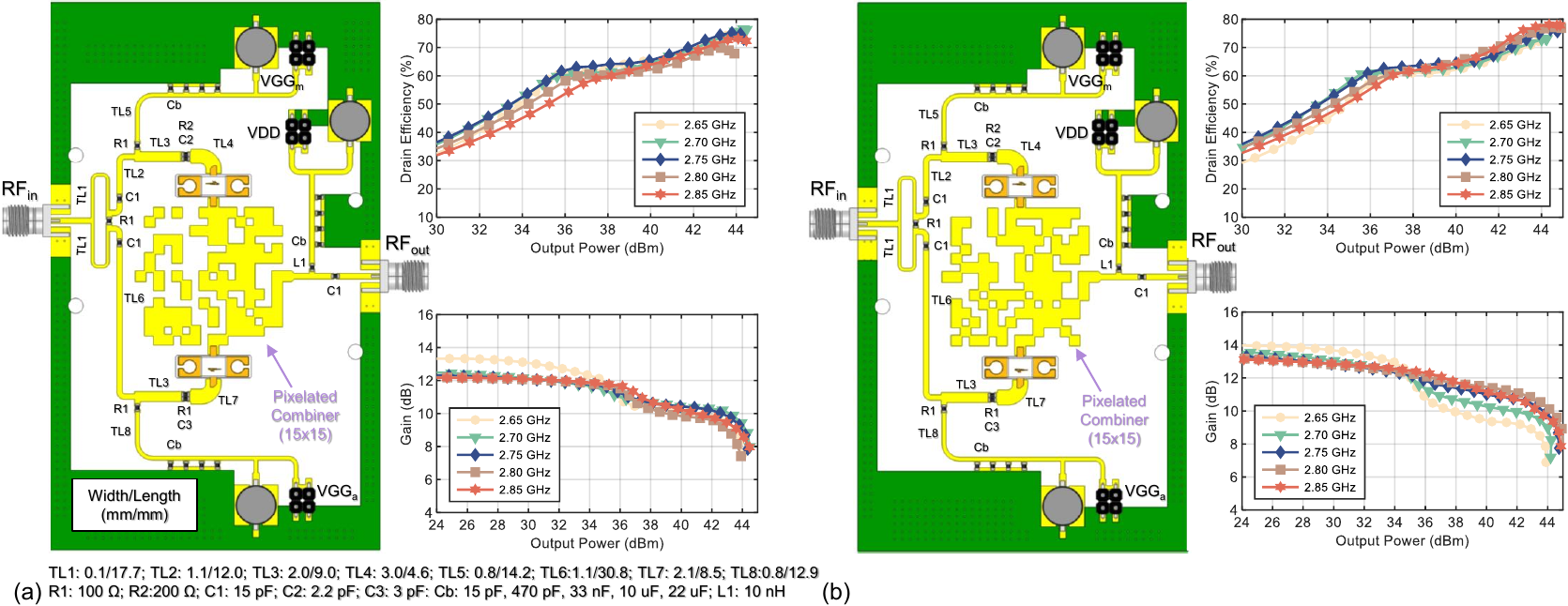}
    \caption{Full circuit schematic of the proposed deep learning-driven pixelated Doherty PAs: a) prototype 1 and (b) prototype 2. The corresponding full EM simulation results show the large-signal performance  in terms of drain efficiency and gain versus output power over the frequency range of $2.65-2.85~\mathrm{GHz}$.}
    \label{fig:schemtaic}
\end{figure*}

\subsection{Pixelated Doherty Combiner Synthesis}
Given the desired two-port impedance matrix derived earlier, we first convert this matrix into its corresponding scattering matrix, which specifies the target S-parameter behavior of the combiner network. To realize a physical implementation matching this target response, we use the GA in combination with the trained deep CNN surrogate model to search the vast pixelated design space efficiently. Unlike traditional approaches that require time-intensive EM simulations for every candidate structure, our deep learning-based method allows the CNN surrogate model to rapidly predict the S-parameters of thousands of pixelated layouts within a second per optimization step. This dramatically accelerates the design process compared to conventional brute-force EM sweeps.

During the optimization, we evaluate each candidate’s predicted S-parameters by calculating a fitness value based on how closely they match the target scattering parameters. The fitness function is defined as
\begin{equation}
F = \frac{1}{e + \epsilon}
\label{eq:fitness}
\end{equation}

\noindent where the error term $e$ is defined as
\begin{equation}
e = \max \left(
\sum \left| \Re\{\hat{S_{i}}\} - \Re\{\Bar{S_{i}}\} \right|, \sum \left| \Im\{\hat{S_{i}}\} - \Im\{\Bar{S_{i}}\} \right|\right)
\label{eq:etot}
\end{equation}

\noindent with $\left|\hat{S_{i}} - \Bar{S_{i}}\right|$ representing the absolute errors between the predicted and target values for both the real and imaginary parts of $S_{11}$, $S_{12}$, and $S_{22}$. A small constant $\epsilon=10^{-5}$ is included in the denominator to ensure numerical stability during fitness calculation.

\begin{figure}[!t]
    \centering   
    \includegraphics[width=0.8\linewidth]{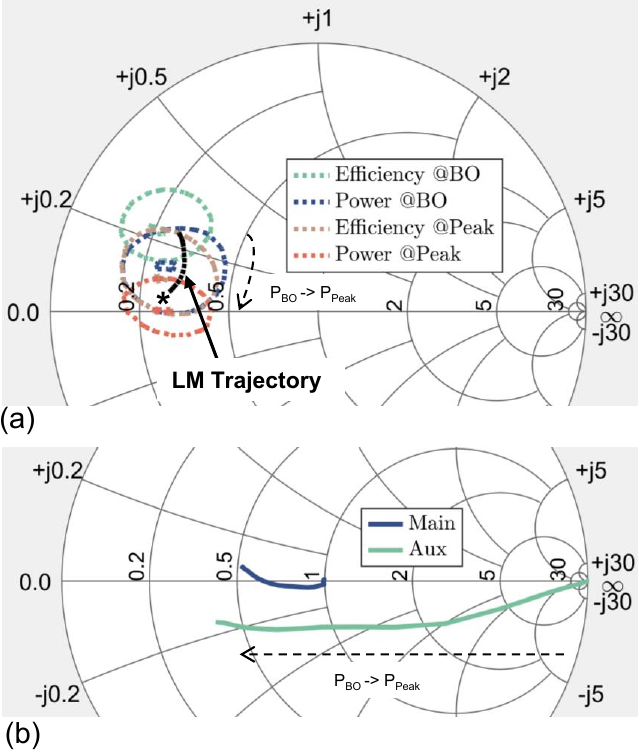}
    \caption{(a) Power‑delivered contours (maximum and -0.3 dB) and efficiency contours (maximum and -3\% of the main amplifier for prototype 1 at peak and back‑off output power, together with its load‑modulation (LM) trajectory. (b) Load‑modulation trajectories of the main and auxiliary amplifiers at the current‑generator plane at the design center frequency.}
    \label{fig:LM_smith}
\end{figure}

Based on this optimization strategy, we synthesize two pixelated Doherty combiner networks, as shown in Fig.~\ref{fig:sim_vs_pred}. The two solutions illustrate the stochastic nature of the GA, revealing that multiple distinct pixelated combiner networks can satisfy the same S-parameter targets for Doherty operation. Note that all three ports are terminated with $50~\Omega$ during the simulations. The agreement observed between the EM‑simulated S‑parameters and the predictions generated by the CNN surrogate model demonstrates reasonable accuracy of the proposed deep learning–assisted design methodology. It is acknowledged that the prediction accuracy is subject to limitations, particularly when modeling S‑parameters with small magnitudes. This agreement could be further improved by expanding the training dataset, enhancing the neural network architecture, and increasing the number of training epochs, given sufficient computational resources. Overall, these results confirm that the proposed approach effectively captures the complex EM interactions within the pixelated structures and enables efficient exploration of a large design space.

\subsection{Simulation Results}
Fig.~\ref{fig:schemtaic} demonstrated the complete circuit schematics of the two fabricated Doherty PA prototypes. EM simulations of all transmission lines are performed using Momentum within Keysight ADS. We use accurate models from Modelithics for all lumped components, and transistor models provided by Macom are employed to ensure reliable simulation results.

The simulated efficiency and gain of the two prototypes versus output power at different operating frequencies are presented in Fig.~\ref{fig:schemtaic}. Excellent efficiency is observed across both saturation and back-off operating points within the targeted frequency band. At the center frequency of $2.75~\mathrm{GHz}$, both prototypes achieve a peak drain efficiency exceeding $78\%$, delivering more than $44.2~\mathrm{dBm}$ saturated output power. Moreover, drain efficiency remains above $60\%$ at the designated $9$-dB back-off point. 

To further illustrate the large‑signal behavior, Fig.~\ref{fig:LM_smith}(a) presents the delivered‑power contours, drain‑efficiency contours, and load‑modulation trajectory of the main amplifier for prototype~1 as a representative example at both peak and back‑off operating points. The delivered‑power contours indicate the maximum output power and the output‑power contour within $0.3$‑dB of this maximum, while the efficiency contours denote the maximum achieved drain efficiency and the corresponding $3\%$ degradation boundary. In addition, the intrinsic load‑modulation trajectory is shown in Fig.~\ref{fig:LM_smith}(b), clearly revealing the characteristic load‑modulation behavior of a Doherty PA. Overall, both prototypes demonstrate consistent large‑signal performance across the entire design bandwidth, thereby validating the robustness and practical viability of the proposed synthesis approach.

\begin{figure}[!t]
  \centering    \includegraphics[width=\linewidth]{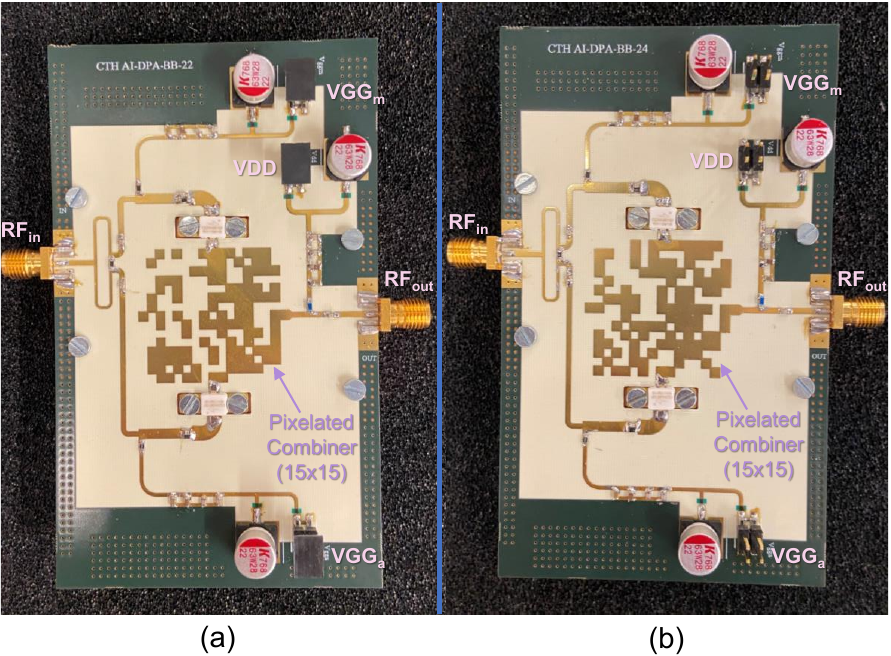}
    \caption{Photograph of the fabricated deep learning-driven Doherty PAs with pixelated output combiners: (a) prototype 1 and (b) prototype 2.}
    \label{fig:prototype}
\end{figure}

\begin{figure*}[!t]
    \centering   
    \includegraphics[width=0.9\linewidth]{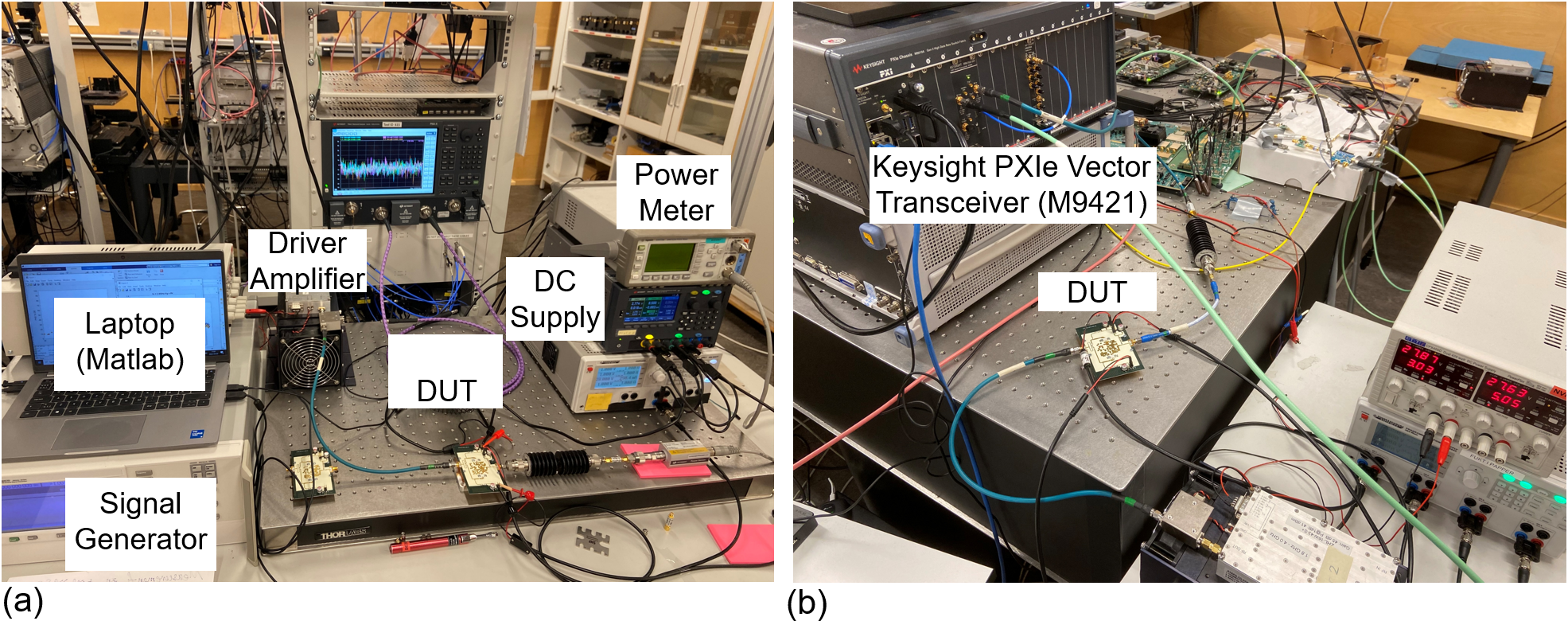}
    \caption{Measurement setup: (a) small‑signal and CW measurements, and (b) modulated‑signal measurements. }
    \label{fig:dpd}
\end{figure*}

\begin{figure}[!t]
  \centering  \includegraphics[width=0.95\linewidth]{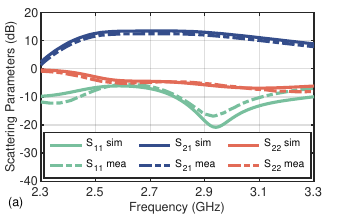}
  \vspace{1mm}
  \includegraphics[width=0.9\linewidth]{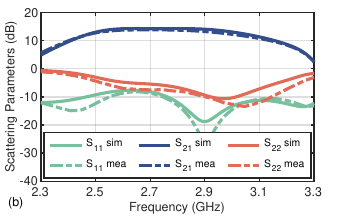}
  \caption{Measured and simulated S-parameters of the fabricated Doherty PAs: (a) prototype 1 and (b) prototype 2.}
  \label{fig:sp}
\end{figure}

\section{Measurement Results}

A photograph of the two fabricated Doherty PA prototypes is shown
in Fig.~\ref{fig:prototype}. The prototype circuits are fabricated on a 20-mil-thick Rogers 4350B substrate, each with a compact size of $60 \times 110~\mathrm{mm^{2}}$. The manufactured printed circuit boards (PCBs) and the employed CG2H40010F GaN HEMT transistors are mounted on brass fixtures, which also serve as heat sinks for thermal management. We perform small-signal, continuous-wave (CW), and modulated-signal measurements to thoroughly characterize the fabricated Doherty prototypes. To ensure consistency, the drain supply voltage is set to $28~\mathrm{V}$ for both prototypes throughout all measurements. Specifically, a quiescent current of $40~\mathrm{mA}$ is set to the main amplifiers, while the gate bias of the auxiliary amplifiers is fixed at $-7~\mathrm{V}$. 

The measurement setup is shown in Fig.~9. Small‑signal measurements are performed using a Keysight PNA‑X network analyzer. For CW measurements, a signal generator provides the CW excitation, and the output power is measured using a power meter. To improve measurement accuracy, a low‑pass filter is inserted at the output to suppress harmonic components and prevent them from entering the power meter. Moreover, a broadband linear driver amplifier supplies sufficient input power to drive the prototype circuits.
For modulated‑signal measurements, a Keysight PXIe vector transceiver (VXT M9421) generates and captures the modulated signals.

\subsection{Small Signal}
The performance of the two Doherty prototypes was first assessed through small‑signal measurements to determine their frequency responses. As shown in Fig.~\ref{fig:sp}, the strong correlation between simulated and measured data confirms proper operation. Within the $2.5–3.1~\mathrm{GHz}$ range, both designs achieve a measured small‑signal gain ($S_{21}$) exceeding 10~dB, while maintaining an input return loss ($S_{11}$) better than -8~dB across the entire band.

\subsection{Continuous Wave}
\begin{figure*}[!t]
  \centering

  \begin{minipage}{0.33\linewidth}
    \centering
    \includegraphics[width=\linewidth]{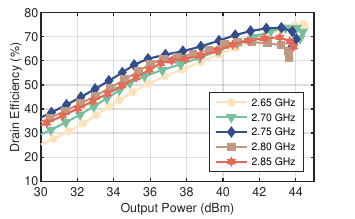}
    \par\vspace{0.5mm}(a)
  \end{minipage}\hfill
  \begin{minipage}{0.33\linewidth}
    \centering
    \includegraphics[width=\linewidth]{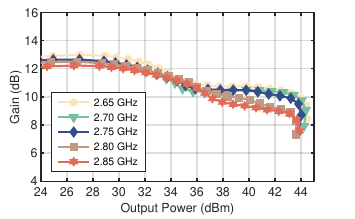}
    \par\vspace{0.5mm}(b)
  \end{minipage}\hfill
  \begin{minipage}{0.33\linewidth}
    \centering
    \includegraphics[width=\linewidth]{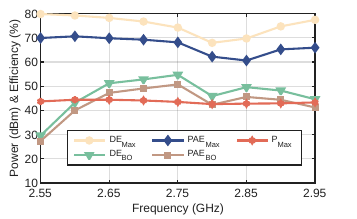}
    \par\vspace{0.5mm}(c)
  \end{minipage}

  \vspace{1.5mm}

  \begin{minipage}{0.33\linewidth}
    \centering
    \includegraphics[width=\linewidth]{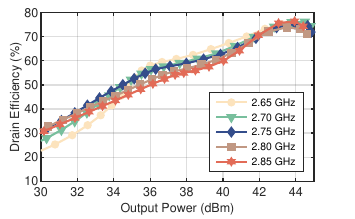}
    \par\vspace{0.5mm}(d)
  \end{minipage}\hfill
  \begin{minipage}{0.33\linewidth}
    \centering
    \includegraphics[width=\linewidth]{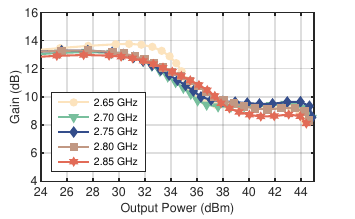}
    \par\vspace{0.5mm}(e)
  \end{minipage}\hfill
  \begin{minipage}{0.33\linewidth}
    \centering
    \includegraphics[width=\linewidth]{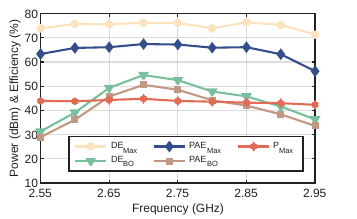}
    \par\vspace{0.5mm}(f)
  \end{minipage}

  \caption{Measured drain efficiency and gain of the Doherty prototype circuits versus output power, along with measured peak and back-off drain efficiency, PAE, and output power versus frequency: (a–c) prototype 1 and (d–f) prototype 2. }
  \label{fig:cw}
\end{figure*}

Fig.~\ref{fig:cw} presents the measured drain efficiency and gain as functions of output power for frequencies spanning $2.65$ to $2.85\mathrm{GHz}$ in $50$-MHz steps. At the center design frequency of $2.75~\mathrm{GHz}$, Doherty prototype 1 achieves a measured peak drain efficiency of $74\%$ with a maximum output power of $44.1~\mathrm{dBm}$, and a measured drain efficiency of $55\%$ at $9$-dB back-off power level. Meanwhile, Doherty prototype 2 reaches a peak drain efficiency of $76\%$ with a peak output power of $44.8~\mathrm{dBm}$, and achieves a measured drain efficiency of $51\%$ at the $9$-dB back-off level. The performance of both prototypes versus frequency is also shown in Fig.~\ref{fig:cw}, where the drain efficiency and power-added efficiency (PAE) at both the peak output power level and the $9$-dB back-off level are presented, along with the peak output power. Across the frequency range from $2.65$ to $2.85~\mathrm{GHz}$, the measured peak and $9$-dB back-off efficiency for both prototypes exceed $68\%$ and $46\%$, respectively.








These results clearly demonstrate that excellent high-efficiency enhancement is achieved around the $9$-dB back-off power level, accompanied by a power gain compression of approximately $3~\mathrm{dB}$. Furthermore, both prototype circuits exhibit consistent CW performance across the entire measured frequency band, confirming the effectiveness and robustness of the proposed design methodology.

\begin{figure*}[!t]
    \centering   
    \includegraphics[width=0.88\linewidth]{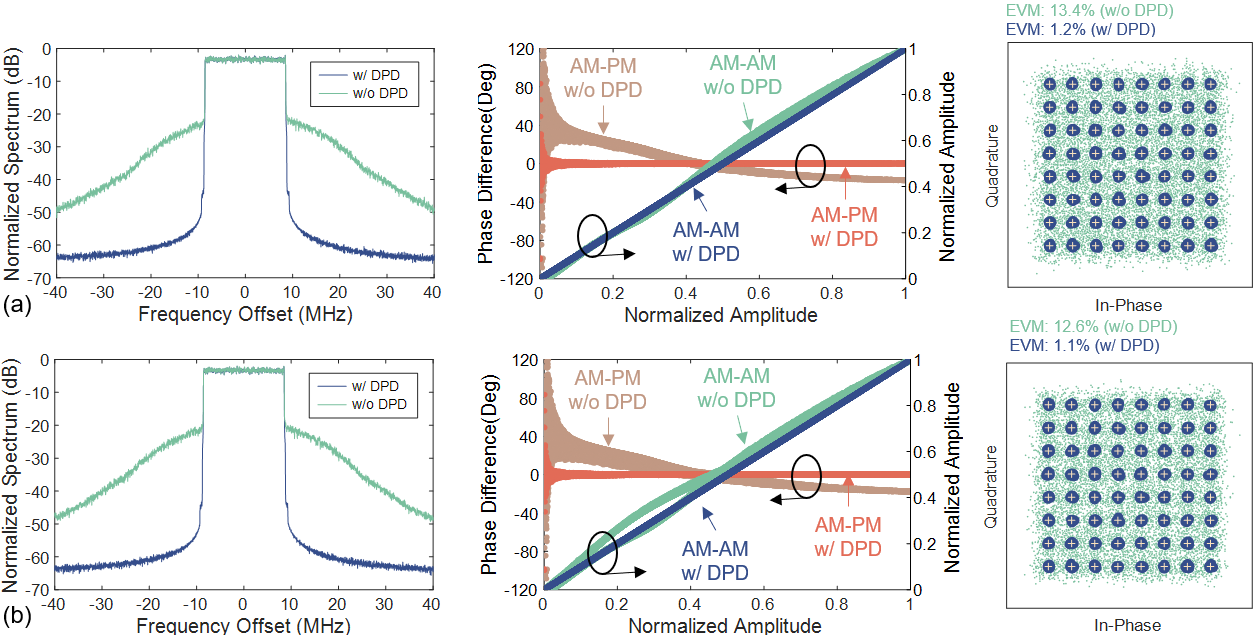}
    \caption{Measured output spectrum, AM–AM/AM–PM characteristics, and constellation diagram under a 20‑MHz, 9‑dB PAPR 5G NR‑like communication signal at 2.75~GHz: (a) Prototype 1 and (b) Prototype 2. }
    \label{fig:dpd}
\end{figure*}

\begin{table*}[t!]
  \centering
  \caption{Benchmarking of State-of-the-Art High-Efficiency PAs.}
  \label{tab:comparison}
  \renewcommand{\arraystretch}{1.1} 
  \begin{tabular}{@{}lccccccccccc@{}}
    \toprule
    Reference & Architecture & \begin{tabular}[c]{@{}c@{}}Freq\\ (GHz)\end{tabular}  & \begin{tabular}[c]{@{}c@{}}$P_\mathrm{max}$\\ (dBm)\end{tabular} & \begin{tabular}[c]{@{}c@{}}OPBO\\ (dB)\end{tabular} & \begin{tabular}[c]{@{}c@{}}DE / PAE\\ @ $P_\mathrm{max}$\\ (\%)\end{tabular} & \begin{tabular}[c]{@{}c@{}}DE / PAE\\ @ OPBO\\ (\%)\end{tabular} & \begin{tabular}[c]{@{}c@{}}Signal\\ BW\\ (MHz)\end{tabular} & \begin{tabular}[c]{@{}c@{}}PAPR\\ (dB)\end{tabular} & \begin{tabular}[c]{@{}c@{}}DE\\ @ $P_\mathrm{ave}$\\ (\%)\end{tabular} & \begin{tabular}[c]{@{}c@{}}ACLR\\ (dBc)\end{tabular} &
    \begin{tabular}[c]{@{}c@{}}Size\\ \end{tabular}\\
    
    \midrule
    \cite{Dohertyblackbox} 2016 & Sym. 2-way DPA & \(\)1.95\(\) & \(\)44.0\(\) & \(\)9\(\) & \(\)68/60\(\) & \(\)52/44\(\) & \(\)20\(\) & \(\)8.5\(\) & \(\)55\(\) & \(-49.0\) & 0.062\\
    \cite{Sym_DPA1} 2023 & Sym. 2-way DPA & 3.20 & $43.2^{*}$ & 6 & $54^{*}/\text{N.A.}$ & $46^{*}/\text{N.A.}$ & 100 & 6.0 & 42.2 & $-49.1$ & 0.037 \\
    \cite{DualIn_DPA2} 2019 & Dual-In Sym. 2-way DPA & \(\)2.00\(\) & \(\)42.9\(\) & \(\)8\(\) & \(\)$70^{*}$/$68^{*}$\(\) & \(\)65/62\(\) & \(\)50\(\) & \(\)9.5\(\) & \(\)53.3\(\) & \(-47.1\) & N.A.\\
    \cite{HarmonicInjection_DPA1} 2022 & Har. Inj. Sym. 2-way DPA & \(\)1.80\(\) & \(\)$43.2^{*}$\(\) & \(\)9\(\) & \(\)$68^{*}$/\text{N.A.}\(\) & \(\)$64^{*}$/\text{N.A.}\(\) & \(\)5\(\) & \(\)9.6\(\) & \text{N.A.} & $-23.5^{**}$ & N.A.\\
    \cite{ClassEF_DPA} 2023 & Class-EF Sym. 2-way DPA & \(\)2.60\(\) & \(\)45.2\(\) & \(\)6\(\) & \(\)76/$75^{*}$\(\) & \(\)74/$74^{*}$\(\) & \text{N.A.} & \text{N.A.} & \text{N.A.} & \text{N.A.} & N.A.\\
    \cite{DualIn_DPA1} 2023 & Dual-In Sym. 2-way DPA & \(\)2.40\(\) & \(\)43.4\(\) & \(\)6\(\) & \(\)70/\text{N.A.}\(\) & \(\)58/\text{N.A.}\(\) & \text{N.A.} & \text{N.A.} & \text{N.A.} & \text{N.A.} & N.A.\\
    \cite{DohertyHan} 2022 & Sym. 3-way DPA & \(\)2.14\(\) & \(\)45.3\(\) & \(\)10\(\) & \(\)69/57\(\) & \(\)55/45\(\) & \(\)20\(\) & \(\)8.5\(\) & \(\)56.6\(\) & \(-49.8\) & 0.091\\
    \cite{RFin_LMBA1} 2022 & RF-Input LMBA & \(\)2.40\(\) & \(\)44.1\(\) & \(\)6\(\) & \(\)54/\text{N.A.}\(\) & \(\)47/\text{N.A.}\(\) & \(\)10\(\) & \(\)8.6\(\) & \(\)44.0\(\) & $-40.5^{**}$ & N.A.\\
    \cite{SLMBA1_Anding} 2020 & RF-Input SLMBA & \(\)3.30\(\) & \(\)$43.2^{*}$\(\) & \(\)10\(\) & \(\)$71^{*}$/$59^{*}$\(\) & \(\)$48^{*}$/$42^{*}$\(\) & \(\)200\(\) & \(\)10.0\(\) & \(\)43.6\(\) & \(-43.9\) & 0.086\\
    \cite{RFin_SLMBA} 2024 & Double-Balanced SLMBA & \(\)2.10\(\) & \(\)40.0\(\) & \(\)15\(\) & \(\)65/\text{N.A.}\(\) & \(\)54/\text{N.A.}\(\) & \(\)20\(\) & \(\)13.0\(\) & \(\)50.2\(\) & $-23.5^{**}$ &0.098\\
    \cite{RFin_CLMA} 2024 & RF-Input CLMA & \(\)3.40\(\) & \(\)42.3\(\) & \(\)6\(\) & \(\)57/\text{N.A.}\(\) & \(\)53/\text{N.A.}\(\) & \(\)20\(\) & \(\)7.0\(\) & \(\)51.0\(\) & \(-51.6\) & N.A.\\
    \midrule
    \multirow{2}{*}{\textbf{This Work}} & \textbf{DL DPA Prototype 1} & \(\mathbf{2.75}\) & \(\mathbf{44.1}\) & \(\mathbf{9}\) & \(\mathbf{74}/\mathbf{68}\) & \(\mathbf{55}/\mathbf{51}\) & \(\mathbf{20}\) & \(\mathbf{9.0}\) & \(\mathbf{53.5}\) & \(\mathbf{-61.1}\) & \(\mathbf{0.06}\)\\
     & \textbf{DL DPA Prototype 2} & \(\mathbf{2.75}\) & \(\mathbf{44.8}\) & \(\mathbf{9}\) & \(\mathbf{76}/\mathbf{69}\) & \(\mathbf{52}/\mathbf{48}\) & \(\mathbf{20}\) & \(\mathbf{9.0}\) & \(\mathbf{51.2}\) & \(\mathbf{-60.8}\) & \(\mathbf{0.06}\)\\
    \bottomrule
  \end{tabular} 
  \\
  \footnotesize N.A. stands for data not available; \textsuperscript{*}Estimated from graph; \textsuperscript{**}No DPD performed;
  \textsuperscript{}Size is defined as the prototype circuit area normalized to the wavelength at the center frequency.
  \end{table*}

\subsection{Modulated Signal}
Modulated‑signal measurements are carried out to validate the proposed Doherty design methodology under realistic wireless communication conditions. Digital predistortion (DPD) is implemented using an iterative learning control (ILC) algorithm~\cite{ILC}, which iteratively refines the input waveform to produce an optimally linearized output from the prototypes. The test signals consisted of 20-MHz 5G new radio (NR)-like waveforms with a PAPR of 9.0~dB. Fig.~\ref{fig:dpd} shows the measured adjacent channel leakage ratio (ACLR) for the two prototypes, which are $-24.8~\mathrm{dBc}$ and $-24.5~\mathrm{dBc}$ at $2.75~\mathrm{GHz}$ without digital predistortion (DPD). When DPD is applied, the ACLR improves significantly to $-60.8~\mathrm{dBc}$ and $-61.1~\mathrm{dBc}$ at the same frequency. After DPD, both prototypes deliver an average output power of about $36~\mathrm{dBm}$ while maintaining an average drain efficiency above $51\%$ and an error vector magnitude (EVM) below $1.2\%$. The measured AM-AM and AM-PM characteristics, both with and without DPD applied, are The measured AM–AM and AM–PM responses, with and without DPD, are also presented, demonstrating that both prototypes achieve excellent linearity once DPD is applied.

\subsection{Performance Comparison}
The performance of the proposed deep learning-driven Doherty PA is summarized in Table~\ref{tab:comparison}, alongside recently reported two-way and three-way Doherty PAs, as well as LMBAs and CLMAs. As can be observed, the designed prototype circuits achieve some of the highest efficiencies at deep power back-off levels, even when compared to other symmetrical two-way Doherty PAs employing dual-input or harmonic injection/tuning techniques. The efficiency is also competitive with that of symmetrical three-way Doherty architectures. In addition, the performance of the proposed design surpasses several RF-input LMBAs and CLMAs that target Doherty-like behavior. Although sequential LMBAs (SLMBAs) demonstrate improved efficiency under extreme power back-off conditions (beyond $10~\mathrm{dB}$), these architectures often experience excessive compression in the main amplifier and demand a larger circuit footprint.

Overall, the proposed deep learning-driven Doherty PAs offer a compelling combination of efficiency, compactness, and design flexibility. The measured linearity results verify that the prototypes satisfy the rigorous linearity and spectral requirements of contemporary wireless communication systems.

\section{Conclusion}
A deep learning-driven inverse design methodology for Doherty PAs with pixelated output combiner networks and extended back-off efficiency is proposed. In this approach, a deep CNN is trained as an EM surrogate model to accurately and efficiently predict the S-parameters of arbitrary pixelated structures. By integrating the surrogate model with a GA optimizer within a black-box Doherty framework, the method enables efficient synthesis of complex three-port combiner networks using fully identical transistors.

To validate the proposed methodology, two Doherty PA prototypes are developed using GaN HEMT transistors. Both prototypes achieve peak drain efficiencies above $74\%$ with output power exceeding $44.1~\mathrm{dBm}$ at the center frequency of $2.75~\mathrm{GHz}$, and maintain efficiencies greater than $52\%$ at the $9$-dB back-off level. Furthermore, under a $20$-MHz 5G NR-like signal with an $9.0$-dB PAPR, the prototypes deliver average efficiencies exceeding $51\%$ with ACLR better than $-60.8~\mathrm{dBc}$ after DPD deployment. These results demonstrate that the proposed deep learning-based methodology effectively enables rapid and accurate design of efficient, linear Doherty PAs suitable for modern wireless communication systems.

\section*{Acknowledgment}
The authors thank Modelithics, Inc., Tampa, FL, USA, for the use of Modelithics models utilized under the University License Program. The authors also acknowledge MACOM for supplying the GaN HEMT transistors used in this work. Furthermore, We would like to thank Prof. María José Madero Ayora of the University of Seville for fruitful discussions and for her support with 5G NR-like communication signal generation and modulated‑signal testing.

\ifCLASSOPTIONcaptionsoff
  \newpage
\fi



%



\bibliographystyle{IEEEtran}
\bibliography{IEEEabrv,Transactions-Bibliography/mybibfile}
%

\vspace{-1 cm}
\begin{IEEEbiography}[{\includegraphics[width=1in,height=1.25in,clip,keepaspectratio]{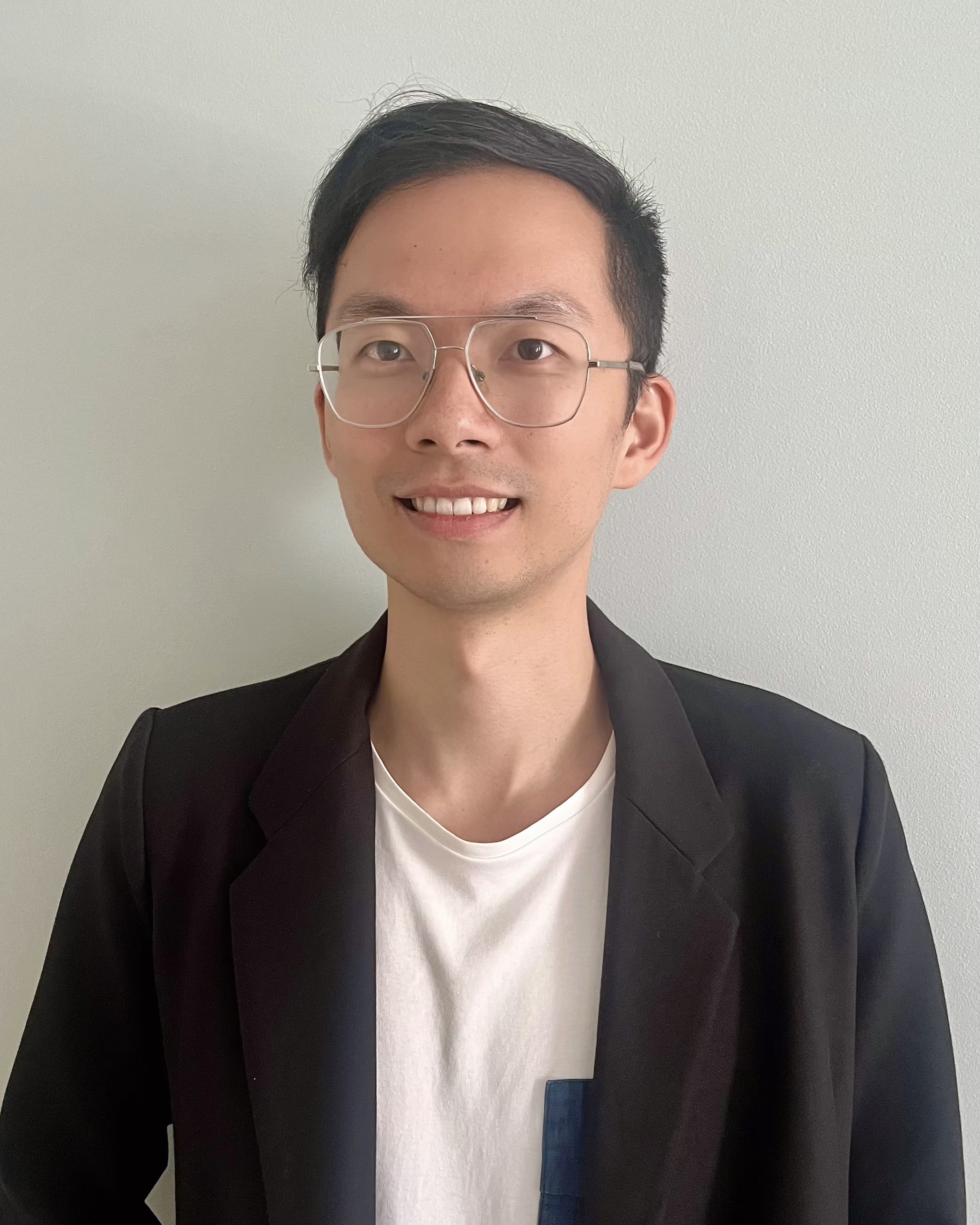}}]{Han~Zhou} (Member, IEEE)
received his B.Sc. degree in space science and astrophysics from the Harbin Institute of Technology, China, in 2016. He received his M.Sc. and Ph.D. degrees in electronic engineering from the Chalmers University of Technology, Sweden in 2018 and 2023, respectively. In 2022, he was a visiting researcher with the IDEAS Group at ETH Zurich, Switzerland. 

He is currently a postdoctoral researcher at Chalmers University of Technology. His research interests include highly efficient, wideband, and linear power amplifier architectures for future wireless transmitters, the design of RF/mm-wave integrated circuits for wireless communication and sensing, and AI-assisted, deep learning-driven design for circuits and systems. 

Dr. Zhou received the IEEE Microwave Theory and Techniques Society (MTT-S) Graduate Fellowship Award in 2023 and the EuMC Young Engineer Prize at the 52nd European Microwave Conference in 2022. He was also awarded a grant from the Ericsson Research Foundation in 2021.

\end{IEEEbiography}
\vspace{-1 cm}
\begin{IEEEbiography}[{\includegraphics[width=1in,height=1.25in,clip,keepaspectratio]{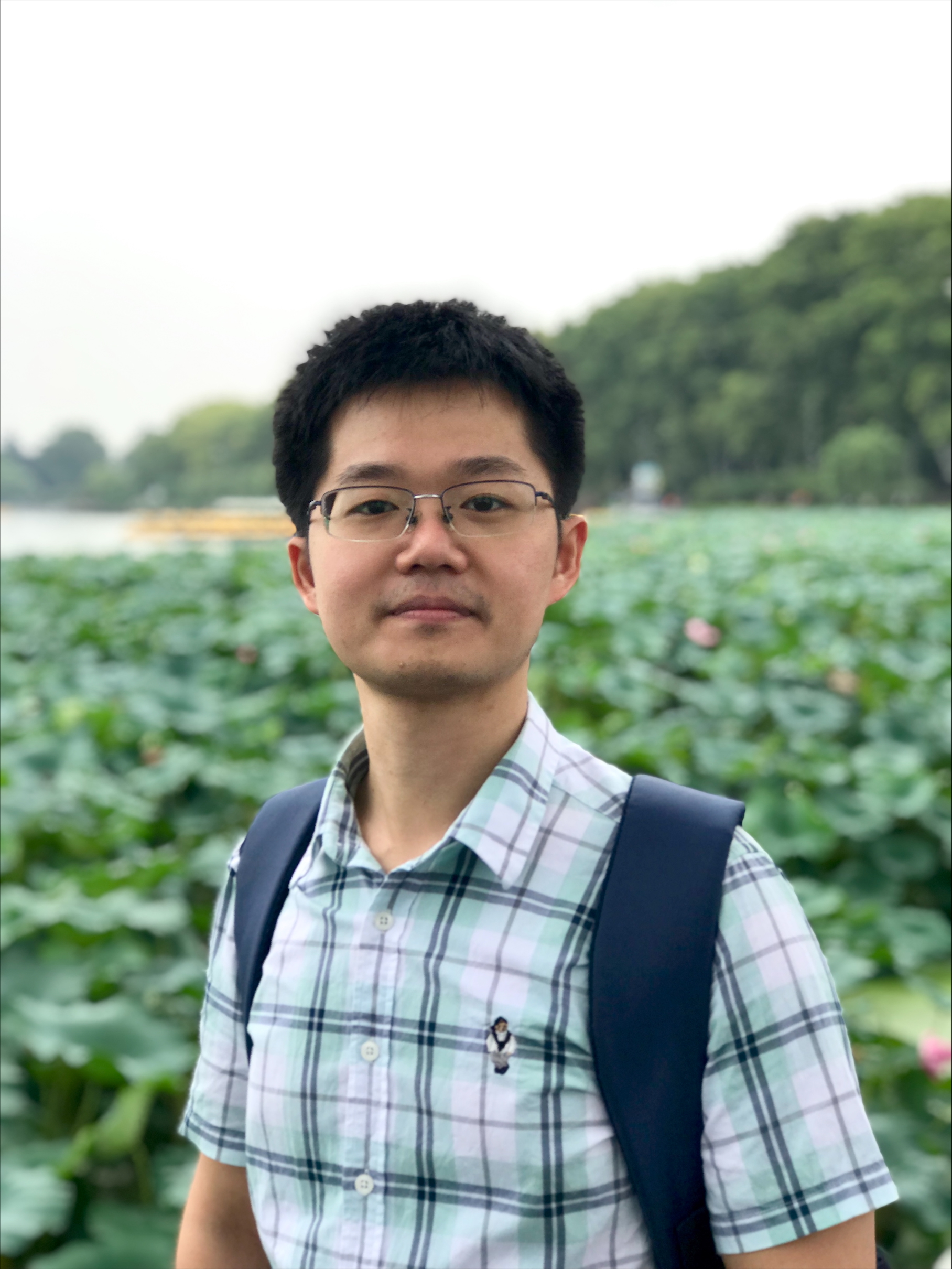}}]{Haojie~Chang}
    received the  Ph.D. degree in  Nanjing University of Science and Technology, China, in 2023. In 2022, he became a visiting researcher in Microwave Electronics Laboratory, Chalmers University of Technology, Sweden and from 2023 he became a post-doc in Chalmers. His current research interests include sub-THz circuits, antenna and packaging design, as well as advanced high-efficiency power amplifier theory and architectures.
\end{IEEEbiography}

\begin{IEEEbiography}[{\includegraphics[width=1in,height=1.25in,clip,keepaspectratio]{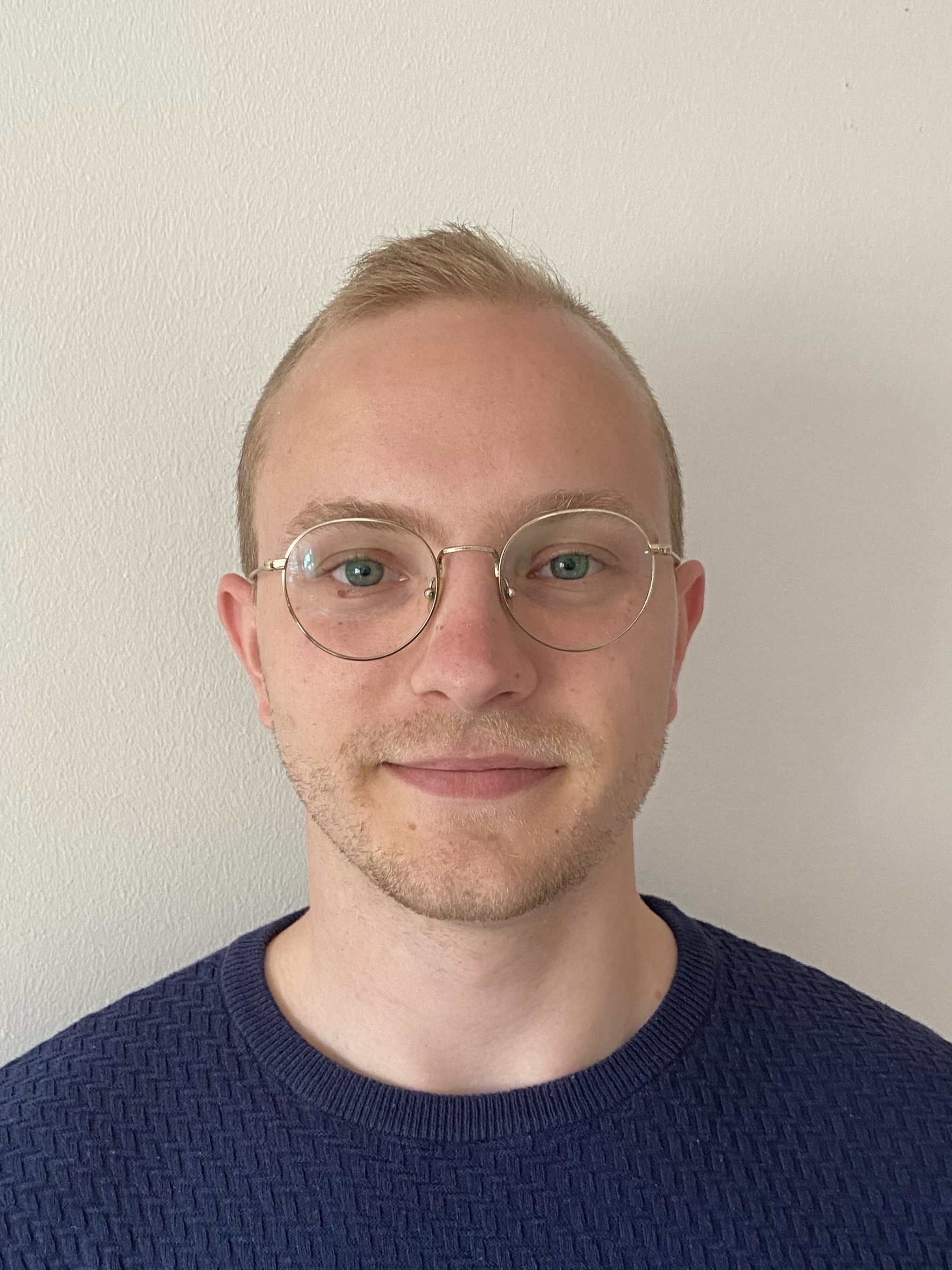}}]{David Widén } (Graduate Student Member, IEEE)
     received the B.Sc. degree in engineering physics from the Chalmers University of Technology, Gothenburg, Sweden, in 2024, where he is currently pursuing the M.Sc. degree in electrical engineering. His research interests include analog circuits and wireless communication.
\end{IEEEbiography}

\end{document}

%% file: mymacros.tex
\usepackage[normalem]{ulem} 
\usepackage{mathtools} 
\usepackage{etoolbox} 
\usepackage{xargs}

\newtoggle{REVIEW}
\toggletrue{REVIEW}  

\def\Put(#1,#2)#3{\leavevmode\makebox(0,0){\put(#1,#2){#3}}}
\newcolumntype{M}[1]{>{\centering\arraybackslash}m{#1}}
\newcolumntype{L}[1]{>{\raggedright\let\newline\\\arraybackslash\hspace{0pt}}m{#1}}
\newcolumntype{C}[1]{>{\centering\let\newline\\\arraybackslash\hspace{0pt}}m{#1}}
\newcolumntype{R}[1]{>{\raggedleft\let\newline\\\arraybackslash\hspace{0pt}}m{#1}}

\makeatletter
\newcommand*{\textlabel}[2]{%
	\ifblank{#2}
	{#1}
	{%
		\def\@currentlabel{#1}
		\leavevmode\phantomsection
		#1\label{#2}
	}
}
\makeatother

\makeatletter
\newcommand{\thickhline}{%
    \noalign {\ifnum 0=`}\fi \hrule height 1pt
    \futurelet \reserved@a \@xhline
}
\newcolumntype{"}{@{\hskip\tabcolsep\vrule width 1pt\hskip\tabcolsep}}
\makeatother

\iftoggle{REVIEW}{
	\def\REVC#1{\textcolor{red}{\it#1}}%
	\newcommandx{\REVN}[2][1=]{\textlabel{\textcolor{blue}{#2}}{#1}}%
	\newcommand{\REVD}[1]{\ifmmode\text{\textcolor{red}{\sout{\ensuremath{#1}}}}\else\textcolor{red}{\sout{#1}}\fi}%
}{
	\newcommandx{\REVN}[2][1=]{#2}%
	\def\REVD#1{}%
	\def\REVC#1{}%
}%